# Response-table-based virtual responding method and instrument and their applications in scanning probe microscopes

Jihui Wang, Qingyou Lu, Xuefeng Cui, Bing Wang

*Hefei National Laboratory for Physical Sciences at Microscale, University of Science and Technology of China (USTC), Hefei, Anhui 230026, People's Republic of China*

We describe a virtual response method and device. It consists of an analog-to-digital converter, a digital-to-analog converter, and a computer and utilizes a searchable response table (RT) pre-stored in the computer to respond to electronic signals. The RT is constructed by measuring the input-output relationship of a real response machine followed by sorting it per input data. To respond, incoming signal is converted to digital data whose position in the RT is then located. The response signal is determined by localized numerical calculation around that position. This method has many advantages: cheap, fast, universal, stable with less noises and errors.

In modern industrial control and scientific research, it is often necessary to respond to certain electronic signals. This can be done by using either a specially designed circuitry or a digital response system with numeric computations. When the response task is complicated, the corresponding circuitry and numeric computations are prone to be complex too. For instance, scanning probe microscopes (SPM) can be realized by either complex circuits[1-4], or complex calculations[5-7] in its digital version.

In this article, we describe a virtual response method and device based on a response table (RT). It gets rid of complex circuits as well as heavy calculations. Its structure is simple and easy to build. We will also show that its response speed is high and its performances in stability, noise, errors and occurrence of malfunction are excellent (in many circumstances, better than its circuitry counterpart). Also, it is cheaper to make and does not require re-build if change to completely different response. It is of general purpose.

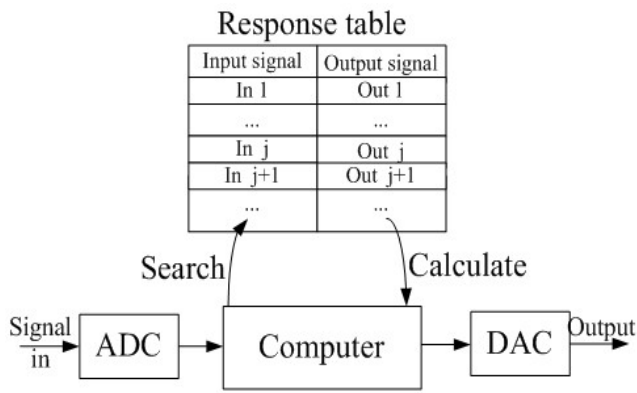

Fig.1 A schematic drawing of response table based virtual responding system

Figure 1 illustrates the basic idea of our response table based virtual responding system. The signal to be responded is amplified by a preamp if necessary. It is then converted to numeric data by an analog-to-digital converter (ADC), which is sent to a computer. There is a response table stored in the computer, which consists of two columns: an input-signal column and its corresponding output-signal column. The response table is sorted by the input-signal column in ascending or descending order. Upon receiving the digitalized input signal to be responded, the computer searches this number in the input-signal column of the response table to locate the smallest interval containing that number. Next, the computer extrapolates the corresponding numeric output signal (numeric responding signal) from the found interval and its corresponding output-signal column. For better accuracy, we can include more data for extrapolation by choosing a larger interval that contains the above found smallest interval.

The algorithm for extrapolation can be interpolation methods, curve fitting methods，or iteration methods. The resulting value can be converted to an analog signal as the output signal (final responding signal) by a digital-to-analog converter (DAC).

The key is how to make the response table. First, we need to find a real and well-performed response machine (which can be called a golden machine). We then statically measure its output signal for each given input. Or we can dynamically measure this response table when it is in real time operation. It is better to make more measurements with finer intervals in regions that need to be responded more frequently. This can improve the accuracy and efficiency. A series of input-output data are attained. Following that, some data processing can be made, for example, sorting by the input signal, deleting fliers, false data, averaging to smoothen the measured response relationship, localized data fitting by segments, or carrying out spectral analysis to get rid of noise/interference spectrum (which may be introduced by the power supply or the electric oscillation in the control circuit, etc). After these processes, a reasonably good response table is obtained.

As an example of applications of the proposed virtual response machine, we can use it to build a scanning tunneling microscope[8] (STM). STM has been improved during the past 25 years since it was invented. It can map out a surface topography of a sample using the tunneling current between the probe and the biased sample. To stabilize the scanning, a feedback control unit is often needed to keep the tunneling current constant. The tunneling current detected by the probe is amplified by a preamplifier and is then sent to the feedback control unit. When the tunneling current is changing during scanning, the feedback control circuits respond to this change by outputting a voltage signal to the Z-piezo, which in turn manipulates tip-sample distance to keep tunneling current constant.

Nowadays, there are many companies and research institutes that offer various designs of feedback control circuits for STMs. Many of them are analog circuitries and very complex. Some are digitized STM employing complicated computation algorithms. We can apply the proposed virtual control system in scanning tunneling microscopes and simulate its effect.

We used an OMICRON UHV-LT-STM as our golden machine. We used sine wave electronic signals generated by Tektronix AFG3000 as the input signals to measure its response table. Sine waves of 5Hz to 80Hz with 30mV amplitude and 100mV DC bias were sent into the feedback control system's input and its output signals were measured at the same time in AC mode.

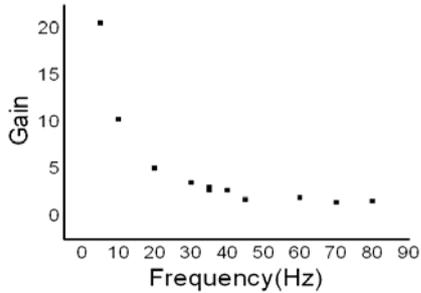

Fig.2 The relationship between the gain (output to input ratio) and the frequency of the input signal

Figure 2 illustrates the relationship between the input signal's frequency and the gain (output to input ratio). The bandwidth is less than 20Hz.

We attained the response table of the feedback control system at a frequency of 20Hz. The input and output signals are both given in Fig. 3. The response has a delay of 9.5ms. After excluding the delay effect, the relationship between each input signal and its corresponding output signal is obtained and illustrated in Fig. 4. After sorting by input signal, a response table was built (called RT1). Then, localized data averaging was carried out, and a smoother response table (called RT2) was built (see Fig. 5).

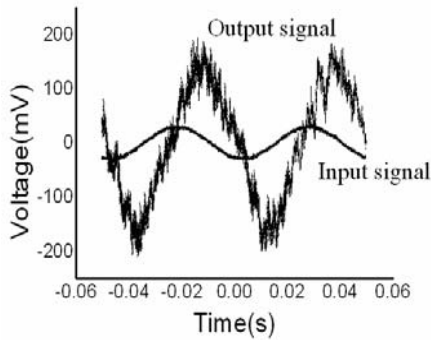

Fig.3 Comparison of input signal and output signal attained at a frequency of 20Hz

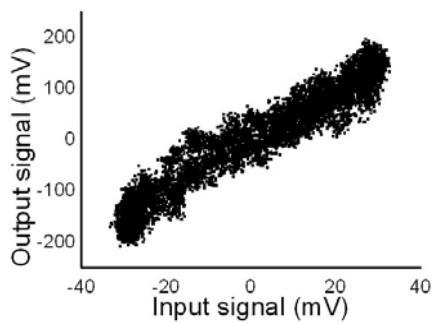

Fig.4 A plot of the directly measured response table after excluding the delay (RT1)

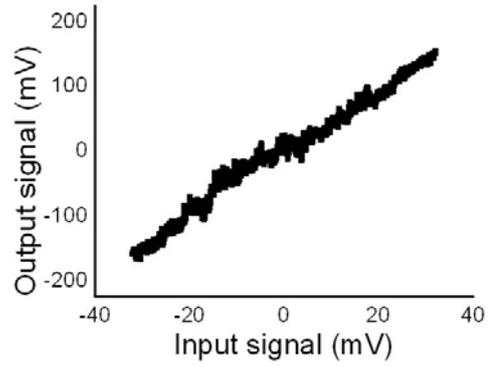

Fig.5 An illustration of the response table (RT2) after the localized data averaging carried out on RT1 in Fig. 4.

We wrote a program to simulate the responding effects using these two tables. As seen in Fig. 3, the feedback control system is very noisy. RT1 was obtained directly from the real feedback control system without any noise disposals. It was used here to represent the performance of the real feedback system. RT2 was built after a noise treatment on RT1. It stands for our virtual response system.

The program accepts a series of testing numbers as input signals and searches (using bisearch) them in the input-signal column of the response table to locate the smallest interval containing each input signal. Next, the program performs a numeric calculation in the found interval using linear interpolation method. Assume the testing value is k, and the smallest interval is $[in_j, in_{j+1}]$ and its counterpart in the output-signal column is $[out_j, out_{j+1}]$, the program uses the follow equation to generate the output signal to respond:

$$Outsignal = \frac{out_j \times (in_{j+1} - k) + out_{j+1} \times (k - in_j)}{in_{j+1} - in_j}$$

We used a sine wave as the testing signal. The sine wave was produced through the equation below:

$$b = \sin a \times 30$$

with $a \in [0, 5\pi]$ and $a$ increasing at a step of 0.00025, where 30 is a normalization factor to make the testing values span the whole response table.

Figure 6 is the comparison of responding effects using the RT1 and RT2 respectively on the testing sine wave. The second curve corresponds to the output signal produced by RT2, which is very closer to the first curve (the input testing sine wave) and better than the third curve (produced by RT1). The second curve is also less noisy than the third. These mean that the responding effect using a smoothened response table is better than a real feedback control unit.

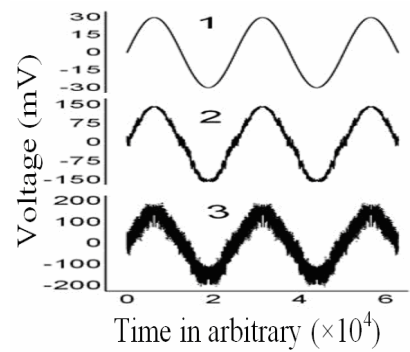

Fig.6 Comparison of responding effects using RT1 and RT2 respectively on the testing sine wave.

As the second test, we used a gray scale map to simulate STM images. We produced a series of STM z values with the equation below:

$$z = (\sin x \times \sin y) \times 28,$$

where, $x \in [0, 3\pi]$, $y \in [0, 3\pi]$, and $x$, $y$ stand for a sample plane (fast and slow scan directions), both increasing at steps of 0.04.

We then added noise signals to this STM image. Random numbers as noise signals were between -2 and 2 generated using Schrage method and were added to each z value as about 7% noise. We used the gray scale to express the z value: the smaller the z value, the brighter the color (Fig. 7a).

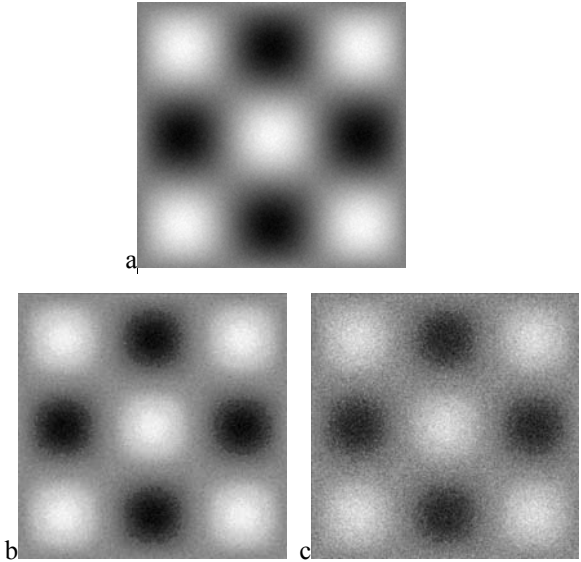

Fig.7 a. The testing gray scale map as input signal (sample signal).    b. The response map produced by RT2.    c. The response map produced by RT1.

Figure 7b presents the simulation result using RT2 to respond to the testing image in Fig. 7a. Fig.7c exhibits the resulting image of simulation by applying RT1 as the response table on the same testing image. Fig. 7b is closer (less noisy) to the testing image than Fig. 7c. It shows higher contrast than Fig. 7c too. These also mean that the responding effect using an improved response table (*i.e.* RT2) is better than the raw feedback control unit (RT1).

The real feedback control unit used in this experiment costs $100k, whereas our virtual response machine costs less than $10k. And when a virtual response machine is built, it can be easily volume-produced because the response table can be copied easily.

The feedback control units of all other SPMs are similar to that of STM. The main difference among SPMs is the different local interactions they sense, which will eventually be converted into voltage signals as the input signals of their feedback control units. Hence, different SPMs can share the same feedback control unit and the proposed virtual response machine is applicable in all types of SPMs.

In this virtual response system, the performance of the computer, ADC, DAC, the efficiency of the program, and the size of the response table, all contribute to its response speed and quality.

The performance indices of ADC mainly include resolution, conversion rate, voltage range and working temperature, etc. Resolution limits the accuracy of a measurement, and is determined by the number of bits. A 20-bit ADC divides the voltage range of the input signal into 1,048,576 discrete levels. With a voltage range of 10 V, the 20-bit ADC can resolve voltage difference as small as 9.5μV. This is a fairly high resolution. There are plenty kinds of ADC from 8 bits to 24 bits in market that can meet different needs. Conversion rate limits the response speed. It can achieve 1 GSPS in current technology. So, using ADC can reduce response time greatly in comparison with complex response circuits. The impacts of voltage range and working temperature can be reduced by using a preamplifier and a temperature controller.

A DAC has similar performance indices as an ADC. It can even achieve a higher conversion rate in the existing technology, doing no harm to the response time.

The impacts of computer speed, program efficiency and the size of response table were also studied. We tested the performance of our program in a computer with Intel CPU 2.13 GHz and memory 504MB. The quantity of input signals for testing was $10^4$, and the size of the response table was from $10^4$ to $10^6$ rows. The relationship between the response time and the size of response table was well fitted by a logarithm function due to bisearch method. When the size of response table hits $10^6$, the total response time for $10^4$ input signals was 46ms. Each response only needs 4.6μs, which is equivalent to a bandwidth better than 200 KHz. The response speed can be adjusted to fit the need of stability[9] conveniently by setting a delay in the response program.

In addition, our proposed virtual response machine is typically less influenced by environment than a specially designed response circuitry. So, it can work more stably and consistently. This device is implemented by very few independent parts, therefore it is very easy to debug provided that the virtual response machine fails to function.

In summary, we have demonstrated a response-table-based virtual responding system with many superiorities especially: 1) it is essentially a type of "hardware copy", which may impact hardware industry as greatly as software copy does on software industry; 2) the "replica" could even be better (performance-wise) than the original; and 3) it is more universal as it can be used in different applications without the necessity of changing the whole system.

We have filed a patent application in China for this technology with application number being 200610085238.4. This work was supported by the 985 project of China.

[1] Raúl C. Munoz, Paolo Villagra, Germán Kremer, Luis Moraga and Guillermo Vidal, Rev. Sci. Instrum. 69, 3259 (1998)

[2] Sang-il Park and C.F. Quate, Rev. Sci. Instrum. 58, 2010 (1987)

[3] E. I. Altman, D. P. DiLella, J. Ibe, K. Lee, and R. J. Colton, Rev. Sci. Instrum. 64, 1239 (1993)

[4] C. Y. Nakakura, V. M. Phanse, G. Zheng, G. Bannon, E. I. Altman and K. P. Lee, Rev. Sci. Instrum. 69, 3251 (1998)

[5] B. A. Morgan and G. W. Stupian, Rev. Sci. Instrum. 62, 3112 (1991)

[6] R. Piner and R. Reifenberger, Rev. Sci. Instrum. 60, 3123 (1989)

[7] C. L. Degen, U. Meier, Q. Lin, A. Hunkeler, and B. H. Meier, Rev. Sci. Instrum. 77, 043707 (2006)

[8] G.Binnig and H.Rohrer, Helv. Phys. Acta. 55, 726

[9] Sang-il Park and C.F. Quate, Rev. Sci. Instrum. 58, 2004 (1987)